\documentclass[review]{elsarticle}

\usepackage{amsmath}
\usepackage{graphicx}
\usepackage{lineno,hyperref}
\modulolinenumbers[5]

\journal{Carbon}

\begin{document}

\begin{frontmatter}

\title{Self-doped graphite nanobelts}


\author[aff_1]{Bruno Cury Camargo\corref{mycorrespondingauthor}}
\cortext[mycorrespondingauthor]{Corresponding author. Tel. +48-22-5532765}
\ead{b.c\_camargo@yahoo.com.br}

\author[aff_2]{Banan El-Kerdi}
\author[aff_3]{A. Alaferdov}
\author[aff_2]{Walter Escoffier}

\address[aff_1]{Faculty of Physics, University of Warsaw, Ul Pasteura 5, 02-093 Warsaw, Poland.}
\address[aff_2]{Laboratoire National des Champs Magnetiques Intenses, CNRS-INSA-UJF-UPS, UPR3228; 143 avenue de Rangueil, F-31400 Toulouse, France.}
\address[aff_3]{Instituto de Fisica Gleb Wattaghin, R. Sergio Buarque de Holanda 777, 13083-859 Campinas, Brasil.}

\begin{abstract}
We report that mechanical deformation of graphite with cavity shock waves introduces a new group of charge carriers, with both effective mass and native concentration one order of magnitude above those found in the pristine material. Their nature, however, remains quasi-2D.  Our results show that defects introduced during mechanical exfoliation have the potential to unlock oscillatory behavior above 50 T in graphite, thus providing a new probe for field-induced electronic phase transitions in the material. 
\end{abstract}


\end{frontmatter}


\section{Introduction}

Although recent years have seen a steady progress in graphene manufacturing \cite{Tatarova2017,Wyss2021}, scalable applications relying on its electronic properties are still scarce. Among other reasons, this happens because substrates severely limit the electronic properties of 2D crystals, and because large-scale graphene manufacturing processes often yields multigraphenes or poor-quality material \cite{Wyss2021, Dorgan2013}. Workarounds these issues often involve expensive and labour-intensive top-bottom encapsulation and exfoliation procedures which, albeit yielding impressive results, are restricted to small-scale laboratory use \cite{Dean2010, Jain2013, Ribeiro2018}. 

A possible alternative to such an issue is to employ multilayer graphenes, or thin graphite flakes in lieu of graphene. Graphite is a quasi-compensated semimetal composed of stacked sheets of graphene. Although single crystals of this material are yet to be achieved and characterized, it is generally accepted that its charge carriers exhibit both low concentration ($\approx 10^{10}$ $\text{cm}^{-2}$) and effective masses ($m \approx 0.05$ $m_e$) \cite{Chung2002}.

Sub-nanometer screening length in this material warrants a better tolerance against impurities and irregularities arising from substrates, while keeping some characteristics of graphene almost intact (such as high electronic mobility, quasi-linear dispersion and electron-hole compensation) \cite{Chung2002, Zhou2006}. This approach, however, comes at the cost of the in-situ control the material's electronic properties, which is a more arduous task in bulk systems. Although feasible on mesoscopic devices properly encapsulated between insulating materials, the necessity of top and bottom gate electrodes makes this process as labour-intensive and non-scalable as in graphene \cite{Miyazaki2008}.

A perhaps naive solution to this dilemma is to directly control the charge carrier concentration in the specimens by modulating graphite's native charge carrier concentration. Due to the high temperatures necessary to synthesize graphite, however, this is not a simple task \cite{Kelly_book}. The presence of dopants usually interferes with the graphitization process, and end up being either eliminated from the final product, or generating graphite of inferior quality \cite{Zhou2010, Stadie2017}. Luckily, the introduction of defects in graphite also has the potential to introduce charge carriers in the material. It has been long demonstrated that vacancies induced by neutron radiation, for example, produce a small charge imbalance in graphite towards holes \cite{Yaguchi1999}. The self-doping imposed by defects in graphite has some advantages over doping achieved with foreign elements. Besides impeding migration (as in the case of mobile interstitial ions \cite{Jeon2020}), the lack of foreign elements guarantees that no contamination would seep out of graphite in various environments \cite{Jeon2020}. It also ensures a better overall chemical compatibility between graphite and its surroundings. 

Here, we take this approach and demonstrate mechanically-treated graphite flakes obtained through liquid phase exfoliation \cite{Alaferdov2018} exhibiting a group of quasi-2D charge carriers with concentration much above those of pristine oriented graphite. Quantum oscillations in this material persist to unusually high magnetic fields, in excess of 50 T. In this field range, they seem to overlap with electronic phase transitions associated to multibody effects in pristine graphite. The latter, denoted by the presence of a magnetic-field-induced high resistance state (HRS), is currently described as a Fermi surface instability triggered exclusively above the the quantum limit.  As we shall see, however, the Landau quantization regime in our system surprisingly coexists with the HRS, making this material an ideal test subject for models aiming at describing the high magnetic field behavior of graphite.

\section{Results and discussion}

\begin{figure}[h]
\includegraphics[width = 8cm]{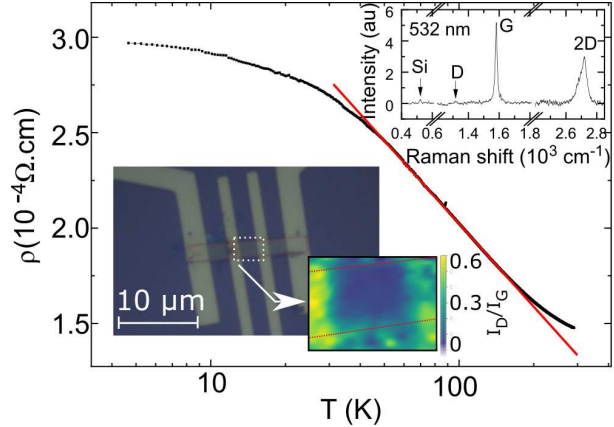}
\caption{Resistance vs. temperature measurements for the sample presented here. The red line is a guide to the eye. The bottom left inset contains a picture of the device, superimposed to a false color map showing the ratio between Raman's D and G peaks in the region between the voltage electrodes. The dashed line indicates the sample boundary, and the resolution of the Raman map is $0.5 \times 0.5$ $\mu \text{m}^2$. The top right inset shows a typical Raman spectrum of the sample. In it, the G, D and 2D peaks of graphite are indicated. Additionally, a Raman peak of the Si substrate at $500 \text{ cm}^{-1}$ is also observed, demonstrating that the measurement samples the entire sample cross section at the LASER spot.}
\label{fig_RxT}
\end{figure}

Samples used here were composed by previously-synthesized narrow graphite belts, few micrometers wide and several micrometers long \cite{Alaferdov2018}. In short, they were obtained by chemically-assisted, liquid-phase exfoliation of natural graphite flakes, followed by a brief annealing treatment at $2950 \text{ }^oC$ for 10 sec. This resulted on small graphite belts with thickness varying between 10 nm and 100 nm, which were then deposited atop 300 nm $\text{SiO}_2$-coated Si substrate, and individually contacted with thermally-evaporated Pd/Au electrodes in a standard 4-probe configuration. A sample picture is shown on Fig. \ref{fig_RxT}. After contacting, the samples were subjected to resistance vs. temperature and magnetic field measurements (R(T) and R(B), respectively) in the temperature range $300$ mK $\leq T \leq$ $300$ K, for $B$ up to 60 T. In total, three samples were measured. They presented the same qualitative results, the most representative of which is shown here.

The samples exhibited insulating-like R(T) curves ($dR/dT<0$), with saturation below 40 K. Such a behavior is typical of disordered graphites \cite{Kelly_book}, with the insulating-like dependency being attributed to a reduction of the carrier mobility with temperature in disordered systems \cite{Klein1961}. 
At intermediate temperatures (between 40 K and 200 K), the resistance followed an $R(T)\approx log (1/T)$ dependence. This functional form suggests a granular material in which the charge carrier concentration increases with T \cite{Efetov2003}. An in-depth analysis of such a behavior in this kind of samples will be published elsewhere \cite{Alaferdov_tbp}.

\begin{figure}[h]
\includegraphics[width = 8cm]{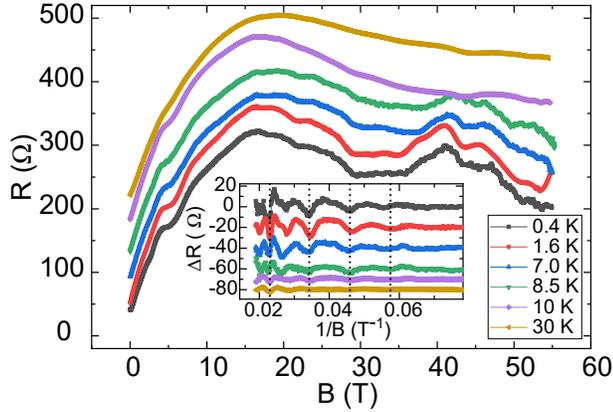}
\caption{Magnetoresistance curves measured at different temperatures. The inset shows their oscillating component $\Delta R$ vs. the inverse magnetic field. The dashed lines are evenly spaced and are a guide to the eye.  A background subtraction attempting to remove the contribution of the HRS was employed (``background 2'' - see fig. \ref{fig_3} and the main text). All curves have been displaced vertically for clarity.}
\label{fig_2}
\end{figure}

\begin{figure}[h]
\includegraphics[width = 8cm]{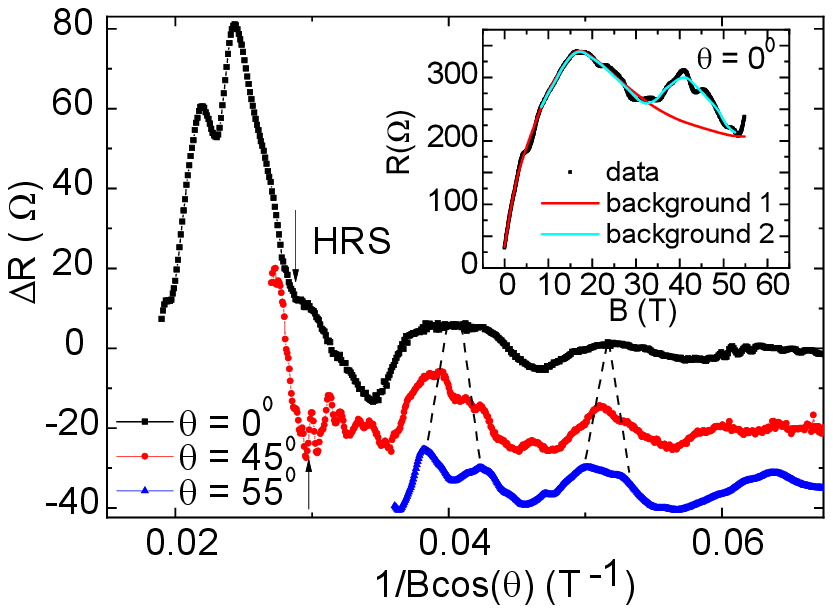}
\caption{Oscillating component of magnetoresistance $\Delta R$ as a function of the reciprocal quantizing magnetic field $1/B\cos(\theta)$. The angle $\theta$ corresponds to the angle between the sample's c-axis and the applied magnetic field. Curves have been displaced vertically for clarity. The contribution due to the magnetoresistance anomaly was not removed (``background 1'' in the inset). Its onset is pointed by black arrows.  Measurements were performed at T = 1.6 K. The dashed lines are guides to the eye, and point regions where a splitting of the maxima are observed with the increase of sample tilting. The inset shows a raw R(T) measurement, obtained at T = 1.6 K and $\theta = 0$ deg. Superimposed to the experimental data are lines used to subtract the non-monotonic MR background while ignoring the HRS (red (darker) line, background 1) and while attempting to remove it (cyan (lighter) line, background 2).} 
\label{fig_3}
\end{figure}

The sample magnetoresistance was also typical of various types of graphite \cite{Soule1958, Camargo2018, Yaguchi2009}. A superlinear $R \propto B^{1.2}$ behavior was observed at low magnetic fields, followed by saturation and a region of negative magnetoresistance above 20 T. Between 35 T and 51 T, a high resistance state (HRS) was observed. The HRS is a characteristic feature of graphite and believed to be associated to a c-axis density-wave transition in the material \cite{Yaguchi2009, Arnold2017}. Superimposed to this non-monotonic MR background, however, an unexpected oscillations could be resolved, which persisted above the temperatures necessary to suppress the HRS. It started around 15 T, and remained up to the highest measured magnetic field. This effect is shown in figs. \ref{fig_2} and \ref{fig_3}, depending on temperature and relative sample orientation to the magnetic field. In order to separate this oscillating component of the magnetoresistance ($\Delta R$) from the rest of the data, two different approaches were taken, as illustrated in fig. \ref{fig_3}. In one of them, a smooth decaying background was considered above 25 T (background 1), whereas on the other a non-monotonic decay was assumed, following the mean of the oscillatory behavior (background 2).  The latter accounted for the removal of both of the region of negative magnetoresistance as well as the HRS feature, at the cost of the accuracy of the quantum oscillations' amplitude for $B> 30$ T. Regardless of the method employed, $\Delta R$ was found periodic with the inverse of the magnetic field (see fig. \ref{fig_2}), akin to the Shubnikov-de-Haas effect (SdH), and apparently exhibited two frequencies:  86 T and its double 172 T. 

This quantum oscillatory behavior scaled with the quantizing magnetic field in tilted magnetoresistance measurements, as shown in fig. \ref{fig_3}. A doubling of the oscillating peak was observed at $B\cos(\theta) = 25$ T ($1/B \approx 0.04$) by increasing $\theta$, which was defined as the angle between the applied field and the sample's c-axis. Such a result suggests that the 172 T  component to the quantum oscillations was associated to the Zeeman splitting of the 86 T component. The scaling of the oscillatory behavior with $B\cos(\theta)$ further indicates that the associated group of charge carriers is confined in-plane, and/or forms a highly anisotropic pocket in graphite's Fermi surface. Considering graphite's band structure \cite{Chung2002}, a 86 T oscillating frequency corresponds to an in-plane carrier concentration of $2.3 \times 10 ^{12} \text{ cm}^{-2}$. 

Their source, however, is unlikely to be associated interstitial and substitutional dopants, as extensive x-ray diffractometry and x-ray photoemission spectroscopy (XPS) did not reveal any sizable quantity of foreign elements (see the SI and ref. \cite{Alaferdov2018}). Back-gating with $V_G=\pm 50$ V yielded no effects in varying the quantum oscillations in the material either, thus suggesting that the phenomenon is associated to bulk charge carriers, rather than associated to quasi-2D charge carriers located at the interface between graphite and the substrate. Raman mapping of the sample, which probed its entire volume (as evidenced by the presence of a Raman line of the underlying Si during measurements on graphite's surface, see fig. \ref{fig_RxT}), also revealed a homogeneous intensity for graphite’s D peak ($\approx 1350 \text{ cm}^{-1}$) throughout the device. The ratio between the D and G peaks was nearly constant, below 1/10.  These results indicate some evenly-distributed disorder on $\text{sp}^2$ bonding across the sample, which did not show a ``hot'' region of defects \cite{Pimenta2007}. 

Curiously, the charge carrier concentration estimated from the 86 T component of the quantum oscillations approximately matched the average density of stacking defects in the material, of $\approx 2\times 10^{19}$ $\text{cm}^{-3}$. These defects, composed by interstitial non-bonded graphene edges and interconnecting planes \cite{Alaferdov2018}, were directly imaged through TEM measurements (see ref. \cite{Alaferdov2018} and the SI). An estimation of their density in our samples was performed by sampling their numbers in cross-sectional TEM images of different nanobelts of the same batch. 

\begin{figure}[h]
\includegraphics[width = 8cm]{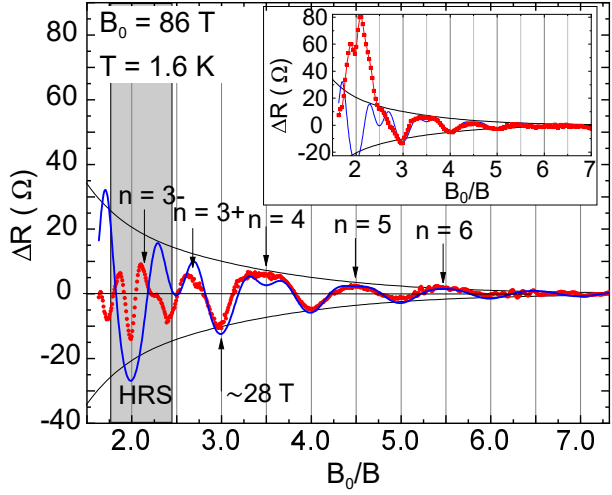}
\caption{Oscillating component of magnetoresistance (red points) at T = 1.6 K, obtained after the removal of a non-monotonic background accounting for the HRS (background 2, see fig. \ref{fig_3}). The blue line is a fit using the L-K model in eq. \ref{eq_LK}, with parameters $m = 0.59$ $m_e$, $B_0 = 86$ T,  and $\tau = 1.35 \times 10^{-13}$ s. The numbers are tentative indexes for the landau levels resolved from the data. The black line is an envelope function of the type $exp(-1/\omega_c \tau_Q).$ The shaded area corresponds to the region where the HRS is observed, outside of which the L-K is satisfactory. The inset shows the same data, with a different background correction (background 1, see fig. \ref{fig_3}).} 
\label{fig_4}
\end{figure}

The  Lifshitz-Kosevich (L-K) model for quantum oscillations in solids describes reasonably well the oscillatory component of the magnetoresistance, down to inverse magnetic fields of $B^{-1} \approx 0.035 \text{ T}^{-1}$. Taking the quantum oscillations frequency at 86 T, this magnetic field corresponds to the emptying of the $n = 3$ LL in the material (see fig. \ref{fig_4}). Below this value of $B^{-1}$, a clear frequency doubling was observed, accompanied by a quantum oscillation amplitude halving. Such a behavior can be attributed as a consequence of a Zeeman splitting $\Delta_s = g\mu_B B$  at a half-integer ratio to the cyclotron energy \cite{Tarasenko2002} in the material ($g$ the gyromagnetic factor and $\mu_B$ the Bohr magneton), thus corroborating our angle-dependent magnetoresistance measurements (fig. \ref{fig_3}). Indeed, the quantum oscillations in this field range were better described by a modified L-K model accounting for a large relative spin splitting \cite{Kunc2015}:

\begin{equation}
\begin{split}
\Delta R \propto \sum_{s=1}^{\infty} (-1)^s \exp{\left[-2 \left(\frac{\pi s}{\omega_c \tau_Q}\right)\right]} \frac{2s\pi^2 k_B T/\hbar\omega_c}{\sinh{2s\pi^2 k_BT/\hbar \omega_c}}  \\ \times \cos{\left(\frac{2s\pi B_0}{B}\right)}\cos{\left(\frac{\pi s \Delta_s}{\hbar \omega_c}\right)},
\end{split} 
\label{eq_LK}
\end{equation}

where $\tau_Q$ is the quantum lifetime of carriers, $\omega_c = eB/m^*$ the cyclotron frequency of the carriers with effective mass $m^*$, $k_B$ the Boltzmann constant, and $1/B_0$ the quantum oscillations frequency. Unfortunately, due to its proximity with the HRS, the background subtraction for magnetic fields $B^{-1}<0.035$ $\text{T}^{-1}$is not as accurate as for the remainder of the experimental window (see fig. \ref{fig_3}), and the amplitude of the quantum oscillations in this region could not be quantitatively evaluated.

Yet, the main oscillatory component of MR in this region allowed the estimation of the spin-splitting parameter $S \equiv (1/2)g(m^*/m_e)=[B_0 \Delta(1/B)]^{-1}$. Here, $\Delta(1/B)$ is the distance between the split peaks of the same Landau level (main sub-peaks, labelled $3+$ and $3-$ in fig. \ref{fig_4}). The electronic Landé g-factor obtained using this relation had a value of $g \approx  2.1 \pm 0.2$, which is - within accuracy - the same value obtained through spin resonance measurements in pristine graphite \cite{Wagoner1960}. We note, however, that the qualitative behavior in this field range is clearly more convoluted, and not fully captured by our simplified approach. For example, each of the two main peaks is apparently composed of two maxima. These fine structures can have different origins, which are beyond the scope of this study.


Outside this field range, the effective electronic masses were extracted from the variation of the amplitude of the quantum oscillations (QO) with temperature (see the SI). They yielded an effective value of ca. $m^* = 0.6\pm0.1$ $m_e$, where $m_e$ is the bare electron mass. The value is about one order of magnitude above the effective mass of carriers commonly found in pristine graphite \cite{Chung2002}. Meanwhile, the quantum scattering rates - extracted from the decay of $\Delta(MR)\times 1/B$ at low temperature - yielded values of $(1.35 \pm 0.4) \times 10^{-13}$ s, which is within the vicinity of $\approx 1.7 \times 10^{-13}$ s previously reported for pristine graphite \cite{Stamenov2005}.

Combined, these results suggest that the new group of charge carriers constitutes a separate band in the graphite samples.  These carriers have a larger mass, a smaller electronic mobility, and one order of magnitude higher charge carrier concentration than those found in the pristine material. Their presence is not usually resolved or discussed in measurements performed on bulk or mesoscopic samples \cite{Camargo2018, Stamenov2005, Fauque2013, Arnold2017, Nakamura1983, Latyshev2010}, being a particularity of our mechanically-treated devices.

An interesting consequence of the results shown here is that the HRS might occur outside the quantum limit in graphite. While this is certainly a possibility (the HRS is triggered concomitantly with the crossing of the $n=3+$ landau subband, therefore occurring outside the quantum limit), we are not able to discard the possibility that small parts of the sample retain their pristine graphite behavior. Therefore, although the triggering of the HRS coincides with the depopulation of the 3+ sublandau band for the carriers discussed here, it is possible that both phenomena are independent. Regardless, the presence of quantum oscillations in this magnetic field range in our samples provide an excellent playground to study subtleties of the magnetic-field-triggered electronic instabilities in graphite.

The new group of charge carriers reported here is not usually seen. On some occasions (e.g. in refs \cite{Timp1983, Taen2018b}) and upon close inspection (e.g.refs \cite{Nakamura1983, Latyshev2010, Taen2018} - see the SI), however, it is possible to resolve sample-dependent oscillating features above 20 T in several devices, both periodic in B and 1/B. When periodic in 1/B, their frequencies vary between 30 T and above 100 T \cite{Nakamura1983, Latyshev2010, Taen2018} (see the SI). Such a behavior, which is seldom addressed, might reflect different degrees of disorder in the material, not always captured by its mosaicity - the preferred parameter to infer sample quality \cite{camargo2016}. A general lack of reports on the high frequency oscillating components on the MR of graphite (or their feeble contribution) might stem, therefore, from the fact that most works regarding the high-magnetic field properties of this material attempt to assess the properties of the HRS, whence utilizing the best-quality available bulk samples. Conversely, here, we utilized shockwave-treated micro-graphite which, albeit possessing high crystallinity, had an inherently higher disorder in comparison with pristine natural flakes \cite{Alaferdov2018}.

\section{Conclusions}
In short, in this report, we demonstrated the presence of a new group of charge carriers in shockwave-treated graphite, with density and effective mass ten times above those found in the pristine material. Our results suggest that the introduced carriers maintain graphite's quantum scattering rates and highly anisotropic band structure, whilst possessing much higher effective mass. Such an approach opens routes towards the study of the electronic correlations occurring in the high magnetic field phase of graphite, as well as for the fabrication of multilayer-graphene-based devices with hard-tuned charge carrier concentration - a desirable feature on carbon-based electronics and analog processing circuits.

\section*{Acknowledgments}
We would like to thank Y. Kopelevich, C. Precker for fruitful discussions and J. Binder for assistance with Raman measurements. This work was supported in part by the National Science Center, Poland, research project no. UMO-2016/23/P/ST3/03514. We acknowledge the support of LNCMI-CNRS, a member of the European Magnetic Field Laboratory (EMFL) under the proposal number TSC09-119. 



%

\end{document}